\documentclass[12pt]{article}

\usepackage[pdftex]{graphicx}
\usepackage[absolute]{textpos}

\usepackage{amssymb}
\usepackage{amsmath}

 \newcommand{\dpar}[2] { \frac{\partial {#1} } {\partial {#2}} }
 \newcommand{\ndpar}[3] { \frac{\partial^{#3} {#1} } {\partial #2 ^{#3} }}

\newcommand{\schr}{Schr\"o\-din\-ger}
\newcommand{\nlse}{nonlinear \schr\ equation}

\newcommand{\ivp}{initial-value problem}

\newcommand{\sech}{\text {\,sech}}

\newcommand{\re}{{\cal R}e}

\newcommand{\Enteros}{\mathbb{Z}}

\title{Classical Forces on Solitons in Finite and Infinite, Nonlinear,
Planar Waveguides}

\author{J. I. Ramos and F. R. Villatoro \\
Departamento de Lenguajes y Ciencias de la Computaci\'on \\
E. T. S. Ingenieros Industriales \\
Universidad de M\'alaga \\
Plaza El Ejido, s/n \\
29013-M\'alaga  \\
SPAIN \\
}

\date{}

\begin{document}

\maketitle

\begin{abstract}

Conservation equations for the mass, linear momentum and energy densities of
solitons propagating in finite, infinite and periodic, nonlinear, planar
waveguides and governed by the \nlse\ are derived. These conservation
equations are used to determine classical force densities which
are compared with those derived by drawing a quantum mechanics analogy
between the propagation of solitons and the motion of a quantum particle in a
nonlinear potential well.

\end{abstract}

\vspace{0.5cm}

KEY TERMS:   Spatial solitons, \nlse, infinite planar waveguides, classical
forces

\section{Introduction}

\label{se:introduction}


Soliton propagation has received a great deal of attention in recent years
because of its ubiquity in many branches of physics, e.g., electrodynamics, fluid
dynamics, plasma physics, etc., and because of its possible applications to
distorsionless signal transmission in ultra-high speed and long-distance
transoceanic telecommunications by optical fibers, soliton lasers, all-optical
soliton switches, nonlinear planar waveguides, nonlinear transmission lines,
etc.~[1]. For example, the envelope of the electric field in a weakly nonlinear,
planar waveguide is governed by the nonlinear Schr\"odinger (NLS) equation
which can be used to determine the forces acting on solitons [2,3].

Based on a quantum mechanics analogy between the NLS equation and the motion of a
quantum particle in a potential well, Bian and Chan [2] and
Ramos and Villatoro~[3]
determined the nonlinear and the diffraction force densities on solitons
propagating in infinite and finite, respectively, weakly nonlinear, planar
waveguides. In particular, Ramos and Villatoro~[3]   determined numerically the
quantum force densities  on  solitons in  finite, weakly
nonlinear, planar waveguides subjected to periodic and homogeneous Dirichlet,
Neumann and Robin boundary conditions at both boundaries of the waveguide.

In this paper, conservation
equations for the mass, linear momentum and energy densities of solitons
governed by the $n$th-dimensional NLS equation are derived. In particular, the
linear momentum equation is used to obtain the classical force
densities which are compared with the quantum force densities derived in
Reference 3. Furthermore, these conservation equations are employed to
assess the changes in linear momentum that a soliton undergoes in a
 finite, weakly nonlinear, planar waveguide as it collides with the
waveguide's boundaries for periodic and homogeneous Dirichlet, Neumann and
Robin boundary conditions.

\section{Conservation Equations and Classical Force Densities on Solitons}

As shown in References  1 and 2, the one-dimensional NLS equation in
dimensionless  Cartesian coordinates can be written as
\begin{equation} \label{eq:nse}
i u_{t} = -u_{xx} - | u |^2 u,  \qquad x\in{\cal{D}}, \quad t \geq 0
\end{equation}
where $\cal D$ is the spatial domain, the subscripts $t$ and $x$ denote partial
differentiation with respect to  time and the spatial coordinate, respectively,
and $u$ denotes the  slowly varying amplitude of the electric field in a weakly,
nonlinear slab waveguide.

Equation~\eqref{eq:nse} can be written in  dimensional form
 by transforming the dependent and independent variables as indicated
in References 2 and 3.

In a recent work [3], the nondimensional NLS
equation was used to draw an analogy between the propagation of a
soliton in a   weakly nonlinear, planar waveguide and the motion of a quantum
particle in a nonlinear potential well. This analogy allowed to determine
both the quantum energy density of and the force densities on solitons
propagating in finite, nonlinear, planar waveguides. The main results obtained
in [3] are repeated here  for completeness and comparison with the classical
mechanics ones derived in this paper, and can be written as
\begin{equation} \label{eq:ene:ld}
e_q(x,t) = \re \left\{ i {u^* u_{t}} \right\}
      = - \re \left\{ u^*u_{xx} + |u|^4 \right\}
      = e_k(x,t) + e_v(x,t),
\end{equation}
where $e$ denotes energy density, the subscripts $q$, $k$ and $v$ stand for
quantum, kinetic and potential, respectively, $\re$ denotes real part, and the
asterisk denotes complex conjugation.

From the quantum energy density, one can easily obtain [3]
\begin{equation} \label{eq:lde:ev2}
f_n(x,t) = - \dpar{}{x} e_v(x,t) =
\dpar{|u|^4}{x} , \qquad f_d(x,t) = - \dpar{}{x} e_k(x,t) = \dpar{}{x} \re
\left\{ u^*u_{xx} \right\},
\end{equation}
where $f$ denotes a quantum force density, and the subscripts $n$ and $d$ stand
for nonlinearity and diffraction, respectively.

Classical conservation equations and classical force densities on solitons may
be obtained from the one-dimensional NLS equation as follows.  The
mass ($m$), linear momentum ($M$) and energy ($E$) densities of the soliton
can be defined as
\begin{equation}
 m = u^* u , \qquad    M = i \left( u_x^* u -
u^* u_x \right) , \qquad   E = |u_x|^2 - \frac{1}{2} |u|^4 .
\end{equation}
Using the NLS equation, i.e., Eq. (1), and the above densities, the following
conservation equations can be easily obtained
\begin{equation} \label{eq:evol1}
\dpar{m}{t} + \dpar{M}{x} = 0 ,
\end{equation}
\begin{equation} \label{eq:evol2}
\dpar{M}{t} = - \dpar{}{x} \left( 4 |u_x|^2 - \ndpar{m}{x}{2} - m^2 \right) =
    - \dpar{}{x} \left( m^2 + 4 E - \ndpar{m}{x}{2} \right) ,
\end{equation}
\begin{equation} \label{eq:evol3}
\dpar{E}{t} = \dpar{}{x} \left( u_x^* \dpar{u}{t} + u_x \dpar{u^*}{t} \right) =
  i \dpar{}{x} \left( u_x^* u_{xx} - u_x {u^*_{xx}} \right) + \dpar{}{x}
(mM).
\end{equation}

It must be noted that the classical energy density can be written as the sum of
potential and kinetic energies using action-angle variables [4].

For initial-value problems, i.e., $-\infty < x < \infty$, $t \ge 0$ and
$|u| \rightarrow 0$ as  $|x| \rightarrow
\infty$, the
integrals  of the left-hand sides of the above three
conservation equations over the whole spatial domain are zero. Therefore, the
total mass or number of particles, the total linear momentum and the total
energy are constant and coincide with the first, second and third invariants of
the initial-value problem of the NLS equation~[4]. The same result applies to
 periodic boundary-value problems. For finite line problems subject to
homogeneous Dirichlet or Neumann boundary conditions at both boundaries, Eq.
(5) indicates that the total mass is conserved, whereas the total linear
momentum (cf. Eq. (6))
is not, in general, conserved.

Similar conservation equations to those derived above
can be obtained in $n$th-dimensional space ($n$=2,3) where the NLS equation can
be written in dimensionless form as
\begin{equation} \label{eq:nse:vect}
i u_{t} = -\nabla^2 u - | u |^2 u,  \qquad {\bf  x}\in{\cal{D}},
  \quad t \geq 0.
\end{equation}
In $n$th-dimensional space, the  mass,
linear momentum and energy densities are defined as
\begin{equation}
m = u^* u , \qquad
{\bf M} = i \left(u \nabla u^* -  u^* \nabla u  \right) , \qquad
E = \nabla u^*  \cdot \nabla u  - \frac{1}{2} |u|^4,
\end{equation}
where the mass and energy densities are scalars and the
momentum is a vector. Using Eq.~\eqref{eq:nse:vect}, one can easily obtain the
following conservation equations
\begin{equation}
\dpar{m}{t} + \nabla {\bf \cdot} {\bf M} = 0,
\end{equation}
\begin{equation}
\dpar{{\bf M}}{t} = - \nabla \left( 4  \nabla u^* {\bf \cdot} \nabla u  -
\nabla^2 m - m^2 \right) = - \nabla \left( m^2 + 4 E - \nabla^2{m} \right)  ,
\end{equation}
\begin{equation}
\dpar{E}{t} = \nabla {\bf \cdot}\left( \dpar{u}{t} \nabla u^* + \dpar{u^*}{t} \nabla u
\right) =
  i \nabla {\bf \cdot} \left( \nabla u^* \nabla^2 u - \nabla u \nabla^2 {u^*}
\right) + \nabla {\bf \cdot} (m{\bf M}),
\end{equation}
which indicate that the total mass, momentum and energy are conserved, i.e.,
they are invariant, for the \ivp\ of the $n$th-dimensional NLS equation if  $|u|
\rightarrow 0$ as $|{\bf x}| \rightarrow \infty$. Note that Eqs. (10)--(12)
reduce to Eqs. (5)--(7) for one-dimensional problems.

From the conservation of linear momentum (cf. Eq. (11)), i.e., from Newton's
second law of classical mechanics, one can easily determine the classical
potential and force densities as
\begin{equation}
V_{cl}= m^2 + 4 E - \nabla^2{m} ,  \qquad  {\bf f}_{cl} = -\nabla{V_{cl}},
\end{equation}
where $V_{cl}$ and ${\bf f}_{cl}$ denote the classical potential and classical
force densities, respectively. Furthermore, one can easily deduce that the
classical potential and force densities (cf. Eq.~(13)) are related to the
quantum energy and force densities (cf. Eqs (2) and (3) in $n$th-dimensional
space) as
\begin{equation}
e_{q} = \frac {1}{4} (V_{cl} - \nabla^2{m} - 3 m^2),
\end{equation}
\begin{equation}
{\bf f}_{cl} =  4 {\bf f}_{q} - \nabla (\nabla^2{m} + 3 m^2) = {\bf f}_{n} + 2
{\bf f}_{d}
   - 2 \nabla  \left( {\nabla u^*{\bf \cdot} \nabla u }  \right).
\end{equation}

It is interesting to note that the above conservation equations for the
mass, linear momentum and energy densities have been derived from
the NLS equation, and have identical form to those of classical fluid
dynamics [5] and electrodynamics~[6,7]. Note that, in
fluid mechanics, the mass conservation equation is also referred to as
Lavoisier's law or continuity equation, while, in electrodynamics, the
conservation of charge is referred to as continuity equation.

  It is also
remarkable that the mass and linear momentum equations derived in this paper
have the same form as those of a lossless, nonlinear transmission line~[8], i.e.,
\begin{equation}  \dpar{v (x,t)}{x} = - L \dpar{I(x,t)}{t}, \qquad
\dpar{I(x,t)}{x} = - C(v) \dpar{v(x,t)}{t}
\end{equation}
where $v$ and $I$ are the voltage and current, respectively, and $L$ and $C(v)$
denote the inductance and capacitance of the nonlinear transmission line per
unit length, respectively. In this case, $m$, $M$ and $V_{cl}$ may be
identified with the capacitance charge, $Q$, $I$ and $v/L$,
respectively. Furthermore, nonlinear LC ladder networks~[9]
may result in evolution equations of the Korteweg-de Vries (KdV) type which have
soliton solutions, and asymptotic methods permit to deduce the NLS equation
from the KdV equation~[10].

\section{Force Densities on Solitons in Infinite, Nonlinear, Planar
Waveguides}

In this section, the classical and quantum
energy and force densities acting on a soliton propagating in an infinite,
 weakly nonlinear, planar waveguide are determined
analytically. The analytical  solution to the initial-value problem of
Eq.~\eqref{eq:nse} can be  written as
\begin{equation}  \label{eq:sol:a}
u(x,t) = A \sech \xi \exp i \eta
\end{equation}
where
\begin{equation} \label{eq:sol:b}
\xi = \frac{A}{\sqrt{2}} ( x - x_0 - c t ) , \qquad
\eta = \frac{1}{2} \left[ c(x-x_0) + \left( A^2 - \frac{c^2}{2}\right) t +
                          \phi_0 \right],
\end{equation}
where $A$, $c$, $x_0$ and $\phi_0$ are the soliton's amplitude, speed, initial
position and intial phase, respectively.

 Using this solution, one can easily
obtain the following quantum densities  \begin{equation} \label{eq:qua:vq}
e_{q} = \left( \frac{c^2}{4} - \frac{A^2}{2} \right) A^2 \sech^2 \xi
\end{equation}
\begin{equation}  \label{eq:qua:fn}
f_n = -2 \sqrt{2} A^5 \sech^4 \xi \tanh \xi
\end{equation}
\begin{equation} \label{eq:qua:fd}
f_d = \frac{A^3}{4} \sech^2 \xi \tanh \xi \left[ \sqrt{2} \left(c^2-2A^2\right)
  + 8 \sqrt{2} A^2 \sech^2 \xi \right]
\end{equation}
\begin{equation} \label{eq:qua:ft}
f_{q} = f_n + f_d = \frac{\sqrt{2}}{4} A^3 \left(c^2-2A^2\right) \sech^2 \xi
   \tanh \xi,
\end{equation}
and the following classical potential and force densities
\begin{equation} \label{eq:cla:vcl}
V_{cl} =  A^2 c^2 \sech^2\xi
\end{equation}
\begin{equation} \label{eq:cla:fcl}
f_{cl} =   \sqrt{2} A^3 c^2 \sech^2\xi \tanh \xi.
\end{equation}

From Eqs. (19) and (23) and Eqs. (22) and (24), the following expression can
be readily obtained
\begin{equation} \label{eq:rel:forces}
\frac {e_{q}}{V_{cl}} = \frac {f_{q}}{f_{cl}} = \frac{1}{4} - \frac{A^2}{2
c^2},
\end{equation}
which indicates that, for the \ivp\ of the NLS equation, the quantum and
classical potential and force densities are proportional to their classical
counterparts. In fact, the magnitude of the quantum force density is 25 per cent
of that of the classical one for $A$=$c$=1 as shown in Figure 1. Note that the
difference in sign between the classical and quantum force densities for
$A$=$c$=1 is due to the definitions of the quantum energy and classical
potential (cf. Eqs. (2) and (13)).

\section{Force Densities on Solitons in Finite, Nonlinear, Planar
Waveguides}

In finite, nonlinear, planar waveguides, Eq. (1) was solved numerically by
means of a second-order accurate, in both space and time, Crank-Nicolson method
subject to
\begin{equation} \label{eq:nse:p}
\ndpar{u}{x}{n}(x,t) = \ndpar{u}{x}{n}(x+2kH,t),
  \qquad \forall n \geq 0, \quad k \in \Enteros,
     \quad x \in {\cal D}, \quad t > 0,
\end{equation}
for periodic waveguides, and
\begin{equation} \label{eq:nse:r}
u(-H,t) + \gamma u_x(-H,t) = 0, \qquad u(H,t) + \gamma u_x(H,t) = 0,
                     \qquad t > 0,
    \end{equation}
 where ${\cal{D}}$=$[-H,H]$, $H=50$, and  the values
$\gamma$=0, $\infty$ and $\gamma \ne 0$ correspond to
homogeneous Dirichlet, Neumann and Robin boundary conditions, respectively,
at both boundaries of the waveguide.

The time step and the grid size used in the calculations shown in the next
section were 0.01 and 0.25, respectively, $\phi_{0}$=$x_{0}$=0 and $A$=$c$=1. The
initial condition corresponded to the exact solution of the \ivp\ of the NLS
equation, i.e., Eq. (17), translated in such a manner that there were not
mathematical incompatibilities between the initial and boundary conditions.

\section{Presentation of Results}

Figures 1--3 show the classical and quantum force densities acting on
solitons which propagate in weakly nonlinear, planar waveguides. The
left side of Figure 1 shows the force densities in an infinite waveguide,
i.e., those corresponding to the \ivp\ of the NLS equation, and indicates
that the magnitude of the classical force density is four times larger than the
quantum one in agreement with the analytical results presented in Section 3.
The left side of Figure 1 also shows the difference in sign between the classical
and quantum force density which is due to the definitions of quantum energy and
classical potential employed in this paper, and the $S$-shape of both force
densities.

The results presented in the right side of Figure 1 correspond to a finite
waveguide subject to periodic boundary conditions. Except for  truncation
errors, the classical and quantum force densities presented in the right side
of Figure 1 coincide with those of the left side of the same figure.
Furthermore,  the linear momentum
 does not change as the soliton interacts with the right boundary in
agreement with the analysis of Section
2.

For a finite waveguide subject to homogeneous Dirichlet boundary conditions
at both boundaries, the left side of Figure 2 indicates that the classical
force densities prior to and after the collision of the soliton with the
right boundary are identical, and that, just during the soliton's interaction
with the boundary, the classical force density undergoes large changes at
the right boundary. The maximum and minimum values of the classical force
density are 0.5637 and -2.2249, respectively, whereas those of the quantum
force density are 0.8059 and -0.7825, respectively. Note that, prior to and
after the interaction of the soliton with the right boundary, the largest
values of the classical and quantum force densities are 0.5443 and 0.1361,
respectively, as shown in the upper left corner of Figure 1.

The left side of Figure 2 shows the classical and quantum force densities on
solitons propagating in finite, planar waveguides subject to homogeneous
Neumann boundary conditions at both boundaries. Due to the large changes in
both the classical and the quantum force densities during the collision of
the soliton with the right boundary, the $S$-shape of these force densities
prior to and after the collision cannot be observed in the right side of Figure
2. The maximum and minimum values of the classical force density are 1.1283 and
-133.2461, respectively, whereas those of the quantum force density are 6.9607
and -44.4007, respectively.

The results presented in Figure 2 clearly indicate that the magnitude
of the classical force density is, in general, much larger and undergoes greater
changes than the quantum one during the interaction of the soliton with the right
boundary.

Since the one-dimensional NLS equation subject to homogeneous Dirichlet or
Neumann boundary conditions is invariant under mirror reflections in $x$, the
interaction of the soliton with the left boundary is expected to be identical
to that with the right one. For this reason, only the interaction of the
soliton with the right boundary was shown in Figure 2.

The results presented in Figure 3 correspond to a finite, planar waveguide
subject to homogeneous Robin boundary conditions at both boundaries and
$\gamma$=1 (cf. Eq. (27)). Since the one-dimensional NLS equation is invariant
under reflections in $x$, while the homogeneous Robin boundary conditions are
not, the interaction of the soliton with the right boundary is expected to be
different from that with the left one as illustrated in the left and right,
respectively, sides of Figure 3. This figure clearly indicates that, due to the
large increase that the classical force density undergoes as the soliton
interacts with the right boundary, the $S$-shape of the classical force density
can be barely seen. The maximum and minimum  values of the classical force
density during the interaction of the soliton with the right boundary are
21.7091 and -1.3303, respectively, whereas those of the quantum force density
are 3.5286 and -0.7605, respectively.

Figure 3 also
indicates that the classical force density undergoes larger variations than the
quantum one and larger variations as the soliton interacts with the right
boundary than when it interacts with the left one. The maximum and minimum
values of the classical force
density during the interaction of the soliton with the left boundary are
0.8590 and -6.1238, respectively, whereas those of the quantum force density
are 0.5064 and -2.3679, respectively.

\section{Conclusions}

The \nlse\ has been used to determine the conservation equations of mass,
linear momentum and energy densities  of solitons propagating in infinite or
finite, weakly nonlinear, planar waveguides. It has been shown that these
conservation equations have the same form as those of classical fluid dynamics
and electrodynamics. It has also been shown that the classical force density
is larger than the quantum one and that it undergoes large changes as the
soliton interacts with boundaries subject to homogeneous boundary conditions.
The largest changes in the classiccal force densities occur in planar
waveguides subject to homogeneous Neumann boundary conditions.

\section*{Acknowledgments}

This research was supported by the Spanish D.G.I.C.Y.T. under Project  no.
PB91--0767. The second author (F.R.V.) has a fellowship from the
Programa Sectorial de Formaci\'on de Profesorado Universitario y Personal
Investigador, Subprograma de Formaci\'on de Investigadores "Promoci\'on General
del Conocimiento", from the Ministerio de Educaci\'on y Ciencia of Spain. The
authors are extremely grateful to Dr. Isabel Prieto Barcia of the Departamento
de Ingenier\'{\i}a de Comunicaciones of the Universidad de M\'alaga for her help
and suggestions.



\newcommand{\bookref}[6]{#1, ``{\em {#2}}." #3, #6, p. #5.}

\newcommand{\paperref}[6]{#1, ``#2," {\em {#3}}, Vol. #4, #6, pp. #5.}

\newcommand{\procref}[7]{#1, ``#2" in ``{\em #3}" edited by #4, #5, pp. #6,
#7.}

\newcommand{\procrefvol}[9]{#1: ``#2" in ``{\em #3}", #4, edited by #5, #6, vol.
#7, pp. #8, #9.}



\newpage
\thispagestyle{empty}
\begin{textblock*}{\paperwidth}(0mm,0mm)
   \noindent\includegraphics[width=\paperwidth,height=\paperheight]{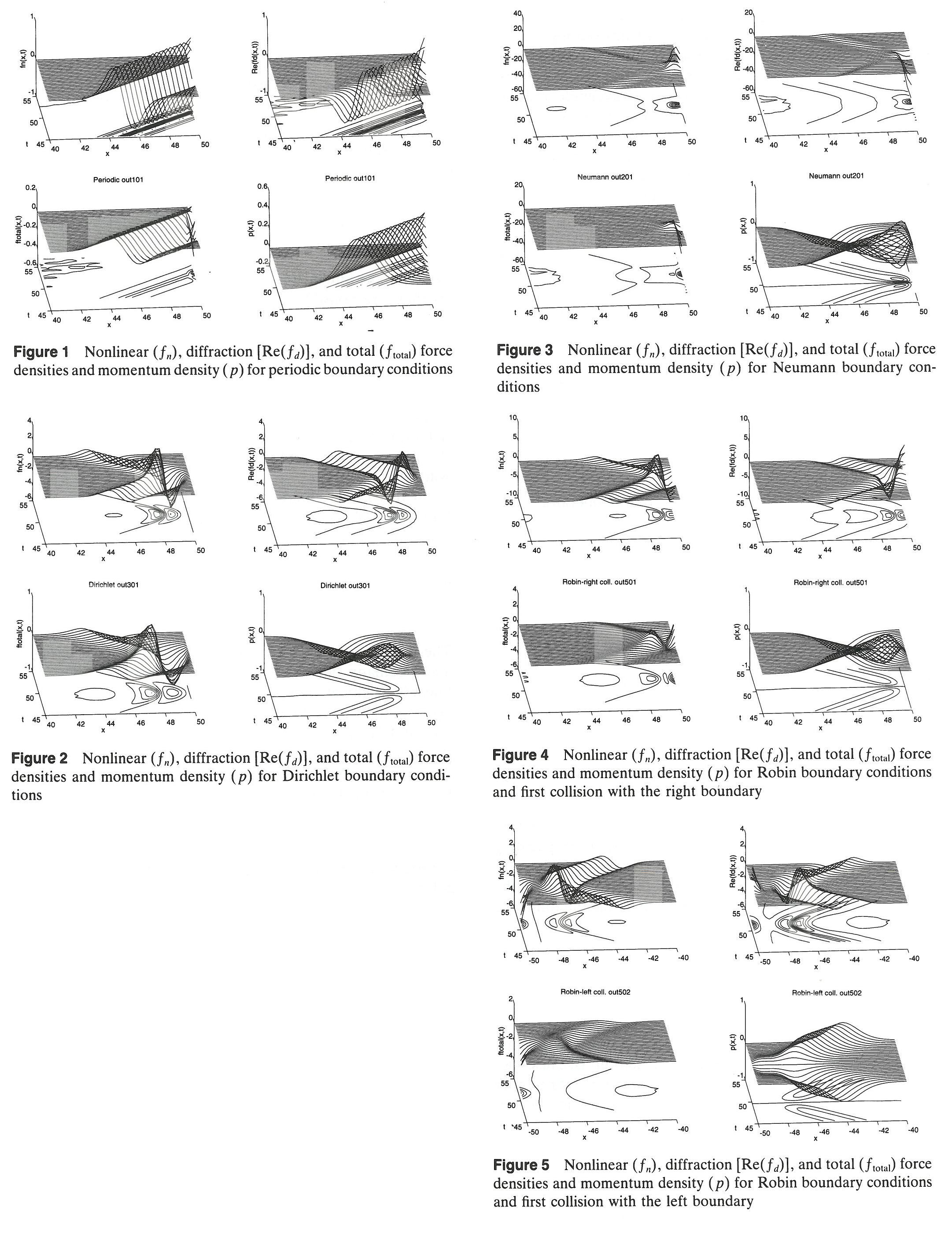}
\end{textblock*}
\mbox{}\newpage

\end{document}